\documentclass[amsmath, pra, twocolumn, showpacs, floatfix]{revtex4-1}
\usepackage{graphicx}
\usepackage{dsfont}

\begin{document}   

\title{Transfer of angular momentum of guided light to an atom with an electric quadrupole transition near an optical nanofiber}
 
\author{Fam Le Kien,$^{1}$ S\'{i}le Nic Chormaic,$^{2}$ and Thomas Busch$^{1}$}

\affiliation{$^1$Quantum Systems Unit, Okinawa Institute of Science and Technology Graduate University, Onna, Okinawa 904-0495, Japan\\
	$^2$Light-Matter Interactions Unit, Okinawa Institute of Science and Technology Graduate University, Onna, Okinawa 904-0495, Japan
}

\date{\today}

\begin{abstract}
We study the transfer of angular momentum of guided photons to
a two-level atom with an electric quadrupole transition near an optical nanofiber. 
We show that the generation of the axial orbital torque of the driving guided field on the atom is governed by the internal-state selection rules for the quadrupole transition and by the angular momentum conservation law with the photon angular momentum given in the Minkowski formulation. We find that the torque depends on the photon angular momentum, the change in the angular momentum of the atomic internal state, and the quadrupole-transition Rabi frequency. 
We calculate numerically the torques for the quadrupole transitions between the sublevel $M=2$ of the hyperfine-structure level $5S_{1/2}F=2$ and the  sublevels $M'=0$, 1, 2, 3, and 4 of the hyperfine-structure level $4D_{5/2}F'=4$ of  a $^{87}$Rb atom. We show that the absolute value of the torque for the higher-order mode HE$_{21}$  is larger than that of the torque for the fundamental mode HE$_{11}$ except for the case $M'-M=2$, where the torque for the mode HE$_{21}$ is vanishing.  	
\end{abstract}

\maketitle

\section{Introduction}
\label{sec:intro}

The angular momentum of light and its transfer to matter have attracted a lot of research attention in recent years \cite{theme issue,Franke-Arnold2017}. 
The transfer of angular momentum of a paraxial light field to particles \cite{Dunlop1995,Dunlop1996,Simpson1997,Dholakia2003}, 
molecules \cite{Babiker2002,Veenendaal2007},
atoms \cite{Babiker1994,Picon2010,Afanasev2013,Lembessis2013}, 
ensembles of cold atoms \cite{Inoue2006,Moretti2009,Nicolas2014,Radwell2015},
and Bose-Einstein condensates \cite{Madison2000,Andersen2006,Ryu2007,Wright2008} has been studied in detail.
For structured light fields, the exchange of angular momentum between an atom and a reflecting surface has been investigated \cite{Donaire2015} and the optical torque on a two-level system near a strongly nonreciprocal medium has been calculated \cite{Monticone2018}. Recently, the transfer of angular momentum from a guided light field of an optical nanofiber \cite{TongNat03,review2016,review2017,review2018} to an atom with an electric dipole transition has been investigated theoretically \cite{chiraltorque}.

Excitations of electric quadrupole transitions of atoms using the guided light fields of optical nanofibers have been experimentally realized \cite{Ray2020}. Unlike electric dipole transitions, electric quadrupole transitions of atoms depend on the gradient of the field amplitude, which can be steep in the case of near fields. In addition, the atomic-internal-state selection rules for quadrupole transitions are more complicated than those for dipole transitions. Consequently, the exchange of angular momentum between a nanofiber-guided light field and an atom for quadrupole transitions  is not simple and deeper insight into the processes involved is desirable. 

The aim of this paper is to study the transfer of angular momentum of nanofiber-guided photons to
a two-level atom via an electric quadrupole transition. 
We show that the generation of the axial orbital torque of the driving guided field on the atom is governed by the internal-state selection rules for the quadrupole transition and by the angular momentum conservation law with the photon angular momentum given in the Minkowski formulation.

The paper is organized as follows. In Sec.~\ref{sec:model}, we describe the model of a two-level atom with an electric quadrupole transition driven by the guided light field of an optical nanofiber.
In Sec.~\ref{sec:torque}, we calculate analytically the azimuthal force and axial orbital torque of the guided light on the atom. In Sec.~\ref{sec:numerical}, we present the results of numerical calculations for the 
torque. Our conclusions are given in Sec.~\ref{sec:summary}.

\section{Model}
\label{sec:model}

We consider a two-level atom with an electric quadrupole transition interacting with the guided light field of an optical nanofiber (see Fig.~\ref{fig1}). We review the descriptions of the atomic electric quadrupole and the guided light field below.

\subsection{Electric quadrupole transition of the atom}

\begin{figure}[tbh]
\begin{center}
  \includegraphics{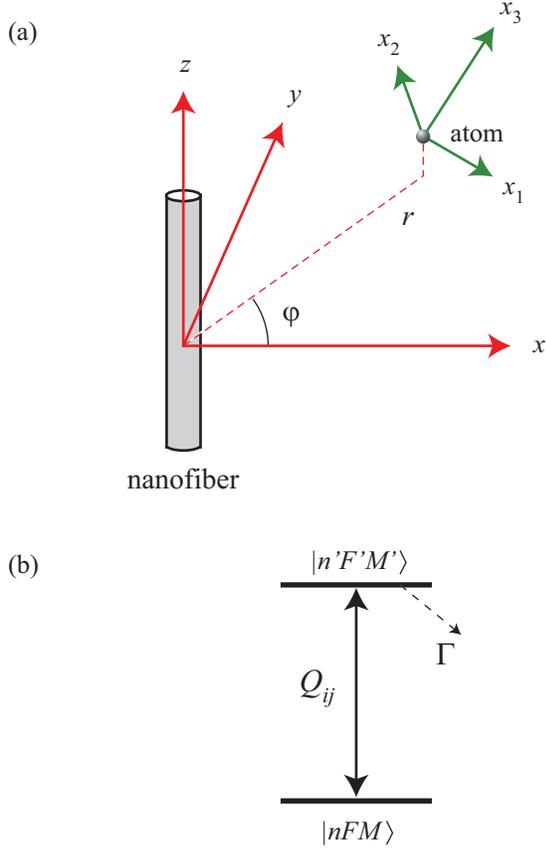}
 \end{center}
\caption{ 
(a) Atom with the local quantization coordinate system $\{x_1,x_2,x_3\}$  
in the vicinity of an optical nanofiber with
the fiber-based Cartesian coordinate system $\{x,y,z\}$ and the corresponding cylindrical coordinate system $\{r,\varphi,z\}$.
(b) Schematic of a two-level atom with an electric quadrupole transition. The upper level $|n'F'M'\rangle$ and the lower level $|nFM\rangle$ of the atom are the magnetic sublevels of an alkali-metal atom. The transition between the two levels is characterized by the electric quadrupole tensor $Q_{ij}$ with $i,j=1,2,3$. The population of the upper level $|n'F'M'\rangle$ may decay with the rate $\Gamma$ into other levels that are not shown in the figure. 
}
\label{fig1}
\end{figure}

We assume that the atom has a single valence electron. To describe the electric quadrupole and the internal state of the atom, we use the local Cartesian coordinate system $\{x_1,x_2,x_3\}$, where the center of mass of the atom is located at the origin $\mathbf{x}=0$ [see Fig.~\ref{fig1}(a)]. The components $Q_{ij}$ with $i,j=1,2,3$ of the electric quadrupole moment tensor of the atom are given as 
\begin{equation}\label{a1}
Q_{ij}=e(3x_ix_j-R^2\delta_{ij}),
\end{equation}
where $x_i$ and $x_j$ are the $i$th and $j$th coordinates of the valence electron and $R=\sqrt{x_1^2+x_2^2+x_3^2}$ is the distance from this electron to the center of mass of the atom. 

Let  $\mathbf{E}$ be the electric component of the optical driving field.
The energy of the electric quadrupole interaction between the atom and the field is
$W=-(1/6)\sum_{ij}Q_{ij}(\partial E_j/\partial x_i)|_{\mathbf{x}=0}$,
where the spatial derivatives of the field components  $E_j$ with respect to the coordinates $x_i$ are calculated at the position $\mathbf{x}=0$ of the center of mass of the atom  \cite{Jackson}.

We assume that the driving field is near to resonance with a quadrupole transition between two atomic internal states, namely the upper state $|e\rangle$ with the  energy $\hbar\omega_e$ and the lower state $|g\rangle$  with the energy $\hbar\omega_g$. 
To be concrete, we consider the quadrupole transition between the magnetic sublevels $|e\rangle=|n'F'M'\rangle$ and $|g\rangle=|nFM\rangle$  of an alkali-metal atom [see Fig.~\ref{fig1}(b)]. 
Here, $n'$ and $n$ denote the principal quantum numbers and also all additional quantum numbers not shown explicitly, 
$F'$ and $F$ are the quantum numbers for the total internal-state angular momenta  of the atom, and $M'$ and $M$ are the magnetic quantum numbers. 
The matrix elements $\langle n'F'M'|Q_{ij}|nFM\rangle$ of the quadrupole tensor operators $Q_{ij}$ are \cite{James1998}
\begin{eqnarray}\label{a5}
\lefteqn{\langle n'F'M'|Q_{ij}|nFM\rangle=3e u_{ij}^{(M'-M)}(-1)^{F'-M'}}\nonumber\\
&&\mbox{}\times
\begin{pmatrix}F' &2 &F \\-M' & M'-M& M\end{pmatrix}
\langle n'F'\|T^{(2)}\|nF\rangle,\qquad
\end{eqnarray}
where the matrices $u_{ij}^{(q)}$ with $q=M'-M=-2,-1,0,1,2$ characterize the tensor structures of the spherical components of $Q_{ij}$ and are given as
\begin{equation}\label{t12}
	\begin{split}
		u_{ij}^{(2)}&=\frac{1}{2}\begin{pmatrix} 1&-i&0\\-i&-1&0\\0&0&0\end{pmatrix},\\
		u_{ij}^{(1)}&=\frac{1}{2}\begin{pmatrix} 0&0&-1\\0&0&i\\-1&i&0\end{pmatrix},\\
		u_{ij}^{(0)}&=\frac{1}{\sqrt6}\begin{pmatrix} -1&0&0\\0&-1&0\\0&0&2\end{pmatrix},\\
		u_{ij}^{(-1)}&=\frac{1}{2}\begin{pmatrix} 0&0&1\\0&0&i\\1&i&0\end{pmatrix},\\
		u_{ij}^{(-2)}&=\frac{1}{2}\begin{pmatrix} 1&i&0\\i&-1&0\\0&0&0\end{pmatrix}.
	\end{split}
\end{equation}
In Eq.~(\ref{a5}), the array in the parentheses is a 3$j$ symbol and the invariant factor $\langle n'F' \| T^{(2)}\|nF \rangle$ is the reduced matrix element of the tensor operators $T_{q}^{(2)}=2(2\pi/15)^{1/2} R^2Y_{2q}(\vartheta,\phi)$. Here, $Y_{lq}$ is a spherical harmonic function of degree $l$ and order $q$,
and $\vartheta$ and $\phi$ are spherical angles in the spherical coordinates $\{R,\vartheta,\phi\}$ associated with the local Cartesian coordinates $\{x_1,x_2,x_3\}$. The matrix $u_{ij}^{(q)}$  represents the tensor structure of  the electric quadrupole operator $Q_{ij}$ for the transition between the magnetic sublevels $|nFM\rangle$ and $|n'F'M'\rangle$ with $M'-M=q$.

We write the electric component of the optical field as $\mathbf{E}=(\boldsymbol{\mathcal{E}}e^{-i\omega t}+\boldsymbol{\mathcal{E}}^\ast e^{i\omega t})/2$,
where $\boldsymbol{\mathcal{E}}$ is the field amplitude and $\omega$ the field frequency. 
The interaction Hamiltonian of the system in the interaction picture and the rotating-wave approximation reads
\begin{equation}\label{a3}
	H_I=-\frac{\hbar}{2}\Omega e^{-i(\omega-\omega_0) t}\sigma_{eg}+\mathrm{H.c.},
\end{equation}
where $\omega_0=\omega_e-\omega_g$ is the atomic transition frequency,  $\sigma_{ge}=|g\rangle\langle e|$ is the atomic transition operator, and
\begin{equation}\label{a4}
	\Omega=\frac{1}{6\hbar}\sum_{ij}\langle e|Q_{ij}|g\rangle\frac{\partial \mathcal{E}_j}{\partial x_i}
\end{equation}
is the Rabi frequency for the quadrupole transition.
We insert Eq.~\eqref{a5} into Eq.~\eqref{a4}. Then, we obtain \cite{James1998}
\begin{equation}\label{a8}
\begin{split}
\Omega&=\frac{e}{2\hbar}(-1)^{F'-M'}
\begin{pmatrix}F' &2 &F \\-M' & M'-M& M\end{pmatrix}\\
&\quad\times
\langle n'F'\|T^{(2)}\|nF\rangle\sum_{ij}u_{ij}^{(M'-M)}\frac{\partial \mathcal{E}_j}{\partial x_i}.
\end{split}
\end{equation}
The electric quadrupole transition selection rules for $F$ and $F'$ and for $M$ and $M'$ are 
$|F'-F|\le 2\le F'+F$ and $|M'-M|\le 2$. For the quantum numbers $J$ and $J'$ of the total electronic angular momenta, the selection rules are $|J'-J|\le 2\le J'+J$. For the quantum numbers $L$ and $L'$ of the orbital electronic angular momenta, the selection rules read $|L'-L|=0,2$ and $L'+L\ge 2$. Note that  the electric dipole transition selection rule for $L$ and $L'$ is $|L'-L|=1$. Consequently, when electric quadrupole transitions are allowed, electric dipole transitions are forbidden.
We also note that the change in the angular momentum of the atomic internal state due to an upward transition is $\hbar(M'-M)$. The selection rules for the quadrupole transitions do not require the equality between this change and the angular momentum of the absorbed photon. For example,  a quadrupole transition between the magnetic sublevels with $|M'-M|=2$
can be caused by a linearly polarized plane-wave light field.

\subsection{Guided light of the optical nanofiber}

We consider the case where the external field interacting with the atom is the guided light field of a nearby vacuum-clad optical nanofiber [see Fig.~\ref{fig1}(a)] \cite{TongNat03,review2016,review2017,review2018}. The fiber is a dielectric cylinder of radius $a$ and refractive index $n_1$ and is surrounded by an infinite background medium of refractive index $n_2$, where $n_2<n_1$. To describe the guided field, we use Cartesian coordinates $\{x,y,z\}$, where $z$ lies along the fiber axis, and also cylindrical coordinates $\{r,\varphi,z\}$, where $r$ and $\varphi$ are the polar coordinates in the cross-sectional plane $xy$. 

We examine the vacuum-clad nanofiber whose radius is small enough so that it can support just the fundamental HE$_{11}$ mode and possibly a few higher-order modes  
in a finite bandwidth around the central frequency $\omega_0=\omega_e-\omega_g$ of the atom \cite{TongNat03,review2016,review2017,review2018}.
The theory of guided modes of cylindrical fibers is described in Ref.~\cite{fiber books} and is summarized and analyzed in detail for nanofibers in Ref.~\cite{highorder}. 

The field amplitude of a quasicircularly polarized hybrid mode $\mathrm{HE}_{lm}$ or EH$_{lm}$ is \cite{fiber books,highorder}
\begin{equation}\label{a20}
\boldsymbol{\mathcal{E}}=(e_r\hat{\mathbf{r}}+pe_\varphi\hat{\boldsymbol{\varphi}}+fe_z\hat{\mathbf{z}})e^{if\beta z+ipl\varphi}.
\end{equation}
Here, $\beta$ with the convention $\beta>0$ is the longitudinal propagation constant determined by the fiber eigenvalue equation, 
$l=1,2,\dots$ and $m=1,2,\dots$ are the azimuthal and radial mode orders, $f=+1$ or $-1$ denotes the forward or backward propagation direction along the fiber axis $z$, and $p=+1$ or $-1$ is the polarization circulation direction index. 
The functions $e_r=e_r(r)$, $e_\varphi=e_\varphi(r)$, and $e_z=e_z(r)$ correspond to the cylindrical components of the 
quasicircularly polarized hybrid mode with $f=+1$ and $p=+1$ and are given in \cite{fiber books,highorder}.

Equation (\ref{a20}) can be used for not only quasicircularly polarized hybrid modes but also transverse electric and magnetic modes.
For the transverse electric and magnetic modes TE$_{0m}$ and TM$_{0m}$, the azimuthal mode order is $l=0$, the mode polarization is single, and the polarization index $p$ can be omitted \cite{fiber books,highorder}. 
The field amplitude of a mode TE$_{0m}$ is given by Eq.~(\ref{a20}) with $l=0$ and $e_r=e_z=0$.
The field amplitude of a mode TM$_{0m}$ is given by Eq.~(\ref{a20}) with $l=0$ and $e_\varphi=0$.

In Eq.~(\ref{a20}), the functions $e_r(r)$, $e_\varphi(r)$, and $e_z(r)$ depend on $r$ but not on $\varphi$ and $z$. Note that the basis unit vectors 
$\hat{\mathbf{r}}=\cos\varphi\,\hat{\mathbf{x}}+\sin\varphi\,\hat{\mathbf{y}}$ and $\hat{\boldsymbol{\varphi}}=-\sin\varphi\,\hat{\mathbf{x}}+\cos\varphi\,\hat{\mathbf{y}}$ depend on $\varphi$. The phase factor $e^{ipl\varphi}$ in Eq.~(\ref{a20}) and the $\varphi$ dependence of the basis vectors $\hat{\mathbf{r}}$ and $\hat{\boldsymbol{\varphi}}$
contribute to the angular momentum of guided light. 
This characteristic has been studied in the Abraham \cite{Partanen2018a,Fam2006,highorder} and Minkowski \cite{Partanen2018a,Fam2006,Bliokh2018,chiraltorque} formulations. It has been shown that the angular momentum per photon in the canonical Minkowski formulation is $\hbar pl$ \cite{Fam2006,Partanen2018a,Bliokh2018,chiraltorque}.

\section{Azimuthal force and axial orbital torque on the atom}
\label{sec:torque}

We assume that the field is in a quasicircularly polarized hybrid $\mathrm{HE}_{lm}$ or EH$_{lm}$ mode, a TE$_{0m}$ mode, or a TM$_{0m}$ mode, that is, the field amplitude is given by Eq.~(\ref{a20}). Let the atom be at a position  $\{x,y,z\}$ in the fiber-based Cartesian coordinates or  $\{r,\varphi,z\}$ in the corresponding cylindrical coordinates. For the local coordinate system $\{x_1,x_2,x_3\}$,
we take $x_1\parallel x$, $x_2\parallel y$, and $x_3\parallel z$. The relation $x_3\parallel z$ means that we use the fiber axis $z$ as the quantization axis for the atomic internal states.

In a semiclassical treatment, the motion of  the center of mass of the atom is governed by the force $\mathbf{F}= -\langle\boldsymbol{\nabla} H_I\rangle$ \cite{coolingbook,dipole force,dipole force 1,dipole force 2} of the driving field.
It follows from the interaction Hamiltonian (\ref{a3}) that the force is 
\begin{equation}\label{c26}
	\mathbf{F}=\frac{\hbar}{2}(\rho_{ge}\boldsymbol{\nabla}\Omega+\rho_{eg}\boldsymbol{\nabla}\Omega^*).             
\end{equation}
Here, we have introduced the notations $\rho_{ij}=\langle i|\rho|j\rangle$ with $i,j=e,g$ for the matrix elements of the density operator $\rho$ for the atomic internal state.

The field amplitude $\boldsymbol{\mathcal{E}}$ [see Eq.~(\ref{a20})] depends on the azimuthal angle $\varphi$ for the position of the atom, and so does the quadrupole-transition Rabi frequency $\Omega$ [see Eq.~(\ref{a8})]. This dependence leads to the azimuthal component 
\begin{equation}\label{c26b}
	F_\varphi=\frac{\hbar}{2r}\bigg(\rho_{ge}\frac{\partial\Omega}{\partial\varphi}
	+\rho_{eg}\frac{\partial\Omega^*}{\partial\varphi}\bigg)             
\end{equation}
of the force $\mathbf{F}$, which is responsible for the rotational motion of the atom around the fiber axis. The axial component of the orbital (center-of-mass-motion) torque on the atom is
\begin{equation}\label{c26a}
	T_z=rF_\varphi=\frac{\hbar}{2}\left(\rho_{ge}\frac{\partial\Omega}{\partial \varphi}
	+\rho_{eg}\frac{\partial\Omega^*}{\partial \varphi}\right).
\end{equation}
It characterizes the rate of the change of the axial component of the orbital (center-of-mass-motion) angular momentum of the atomic system.

Equation (\ref{a8}) indicates that the Rabi frequency $\Omega$ depends on the sum
$\sum_{ij}u_{ij}^{(M'-M)}(\partial \mathcal{E}_j/\partial x_i)$ with  $M'-M=q=0,\pm1,\pm2$. With the help of
expressions (\ref{t12}) and (\ref{a20}), we find 
\begin{equation}\label{a23a}
	\sum_{ij}u_{ij}^{(q)}\frac{\partial \mathcal{E}_j}{\partial x_i}=
V_q(r) e^{if\beta z+i(pl-q)\varphi},
\end{equation}
where
\begin{eqnarray}\label{a23}
	V_0(r)&=&-\frac{1}{\sqrt{6}}
	\Big(e'_r+\frac{e_r}{r}+\frac{il}{r}e_\varphi-2i\beta e_z\Big),
	\nonumber\\
	V_{\pm1}(r)&=&
	\mp\frac{1}{2}f\Big[i\beta(e_r \mp ip e_\varphi)+e'_z\pm\frac{pl}{r}e_z\Big],
	\nonumber\\
	V_{\pm2}(r)&=&
	\frac{1}{2}\Big[e'_r \mp ip e'_\varphi-\frac{1 \mp pl}{r}(e_r \mp ip e_\varphi)\Big].
\end{eqnarray}
Here, the notations $e'_{r,\varphi,z}=\partial e_{r,\varphi,z}/\partial r$ have been introduced. 
Equations (\ref{a23}) show that the factors $V_q(r)$ depend on $r$ but not on $\varphi$ and $z$. Then, it follows from
Eq.~(\ref{a23a}) that the dependence of the sum $\sum_{ij}u_{ij}^{(M'-M)}(\partial \mathcal{E}_j/\partial x_i)$ on $\varphi$ is given by the phase factor $e^{i(pl-M'+M)\varphi}$, and hence so is the dependence of the Rabi frequency $\Omega$ on $\varphi$ [see Eq.~(\ref{a8})]. This leads to
\begin{equation}\label{b9b}
	\frac{\partial\Omega}{\partial\varphi}=i(pl-M'+M)\Omega. 
\end{equation}
Then, Eq.~\eqref{c26a} yields
\begin{equation}\label{b9g}
	T_z=\frac{i\hbar}{2}(pl-M'+M)(\rho_{ge}\Omega-\rho_{eg}\Omega^*).
\end{equation}
Note that the time evolution equation for the population $\rho_{ee}$ of the atomic upper state  $|e\rangle$ reads   \cite{coolingbook}
\begin{equation}\label{b9c}
	\dot{\rho}_{ee}=\frac{i}{2}(\rho_{ge}\Omega-\rho_{eg}\Omega^*)-\Gamma{\rho}_{ee}.
\end{equation}
Here, $\Gamma$ is the total rate of decay of the excited state $|e\rangle$, which includes not only the rate of decay to the ground state
$|g\rangle$ but also the rate of decay to other states.
Hence, the axial component $T_z$ of the orbital torque of the driving field on the atom satisfies the equation
\begin{equation}\label{b7}
	T_z=\hbar(pl-M'+M)(\Gamma\rho_{ee}+\dot{\rho}_{ee}).
\end{equation}

Equation \eqref{b7} is an expression of the angular momentum conservation law. It governs the exchange of angular momentum between the guided driving field and the two-level atom with a quadrupole transition. According to this equation, the magnitude and sign of the axial torque $T_z$ depend on the factor $\hbar(pl-M'+M)$, where $\hbar pl$ stands for the canonical angular momentum  of a photon in the guided driving field in the Minkowski formulation \cite{Fam2006,Partanen2018a,Bliokh2018,chiraltorque},
and $\hbar(M'-M)$ stands for the change of the total internal-state (spin) angular momentum of the atom due to an upward transition. The factor $\Gamma\rho_{ee}+\dot{\rho}_{ee}$ on the right-hand side of Eq.~\eqref{b7} is the absorption rate, where $\Gamma\rho_{ee}$ is the scattering rate and $\dot{\rho}_{ee}$ is the atomic excitation rate \cite{coolingbook}.  Equation \eqref{b7} indicates that the angular momentum of absorbed guided photons is converted into the orbital and spin angular momenta of the atomic system. In addition, we see that the angular momentum of the guided photon imparted on an atom near a nanofiber is of the Minkowski form. This is in agreement with the results of Refs.~ 
\cite{Partanen2018a,Jones1978,Pritchard2005,Milonni2005,Milonni2010,Anzetta2018a,Anzetta2018b,Partanen2018b}.

According to Eqs.~(\ref{b9g}) and (\ref{b7}), the torque $T_z$ is vanishing when $pl=M'-M$, that is, when the Minkowski angular momentum $\hbar pl$ of an absorbed guided photon is equal to the change $\hbar (M'-M)$ of the angular momentum of the atomic internal state. When $pl\not=M'-M$, a nonzero axial torque $T_z$ can appear. It is interesting to note that Eqs.~(\ref{b9g}) and (\ref{b7}) are in agreement with the results for the torque of guided light on a two-level atom with an electric dipole transition \cite{chiraltorque}.

Equation (\ref{b9g}) shows that the torque $T_z$ depends on the quadrupole-transition Rabi frequency $\Omega$ and the atomic coherence $\rho_{ge}$.  
The time evolution of $\rho_{ge}$ is governed by the equation  \cite{coolingbook}
\begin{equation}\label{b9c}
\dot{\rho}_{ge}=\frac{i}{2}\Omega^*(\rho_{ee}-\rho_{gg})-(i\Delta+\Gamma/2)
{\rho}_{ge},
\end{equation}
where $\Delta=\omega-\omega_0$ is the detuning of the field frequency. In the weak-field limit, where the condition $|\Omega|\ll \Gamma$ is satisfied, we can use the approximations $\rho_{ee}\cong0$, $\rho_{gg}\cong1$, and $\dot{\rho}_{ge}\cong0$. In this case, Eq.~(\ref{b9c}) yields $\rho_{ge}=\Omega^*/(i\Gamma-2\Delta)$. Inserting this expression for $\rho_{ge}$ into Eq.~(\ref{b9g}), we obtain
\begin{equation}\label{a24}
T_z=\hbar(pl-M'+M) \frac{|\Omega|^2}{4\Delta^2+\Gamma^2}\Gamma.
\end{equation}
It is clear that if $\Omega\not=0$ then we have $T_z<0$ or  $T_z>0$ for $pl<M'-M$ or $pl>M'-M$, respectively.

Like the absorption of light, the scattering of light also changes the angular momentum of the atom.
The description of the scattering is beyond the framework of the model Hamiltonian (\ref{a3}).  
Similar to the case of atoms with dipole transitions \cite{chiraltorque}, the axial orbital torque of scattering of light due to the quadrupole transition between the levels $M'$ and $M$ is found to be $T_z^{\mathrm{(scatt)}}=\rho_{ee}T_z^{\mathrm{(spon)}}$, where $T_z^{\mathrm{(spon)}}$ is the axial orbital torque due to quadrupole spontaneous emission and is given as
\begin{equation}\label{b15}
T_z^{\mathrm{(spon)}}=\hbar(M'-M)\Gamma_{M'M}-\hbar\sum_{\mu_0}pl\gamma_{\mu_0}-\hbar\sum_{\nu_0}l\gamma_{\nu_0}.
\end{equation}
In Eq.~(\ref{b15}), $\gamma_{\mu_0}$ and $\gamma_{\nu_0}$ are the rates of quadrupole spontaneous emission into the resonant guided mode $\mu_0=(\omega_0 N f p)$ and the resonant radiation mode $\nu_0=(\omega_0 \beta l p)$, and $\Gamma_{M'M}=\sum_{\mu_0}\gamma_{\mu_0}+\sum_{\nu_0}\gamma_{\nu_0}$ is the total quadrupole decay rate. The index $\mu=(\omega N f p)$ labels the guided modes, where $N=\mathrm{HE}_{lm}$, EH$_{lm}$, TE$_{0m}$, or TM$_{0m}$ is the mode type. Here, $l=1,2,\dots$ for HE and EH modes or $0$ for TE and TM modes and $m=1,2,\dots$ are the azimuthal and radial mode orders \cite{fiber books}. The index $\nu=(\omega \beta l p)$ labels the radiation modes, where   
$l=0,\pm1,\pm2,\dots$ is the mode order and $p=+,-$ is the mode polarization index  \cite{fiber books}.

\section{Numerical results}
\label{sec:numerical}

In this section, we present the results of numerical calculations for the axial
torque $T_z$ of the guided light field on an atom with an electric quadrupole transition. 
As an example, we study the electric quadrupole transition between the ground state $5S_{1/2}$ 
and the excited state $4D_{5/2}$ of a $^{87}$Rb atom.
For this transition, we have $L=0$, $J=1/2$, $L'=2$, $J'=5/2$, and $I=3/2$.
The wavelength of the transition is $\lambda_0=516.5$ nm. It is known that the experimentally measured oscillator strength
of the quadrupole transition $5S_{1/2}\to 4D_{5/2}$ in free space is 
$f_{JJ'}^{(0)}=8.06\times 10^{-7}$ \cite{Nilsen1978}. The reduced quadrupole matrix element $\langle n'J'\|T^{(2)}\|nJ\rangle$ is calculated from 
$f_{JJ'}^{(0)}$ by using the relation  \cite{James1998,Tojo2005b}  
\begin{equation}\label{r5}
f_{JJ'}^{(0)}=\frac{m_e\omega_0^3}{20\hbar c^2}\frac{|\langle n'J'\|T^{(2)}\|nJ\rangle|^2}{2J+1},
\end{equation}
where $m_{e}$ is the mass of an electron.
In our numerical calculations, we assume that the driving field is at exact resonance with the atom ($\omega=\omega_0$).

\begin{figure}[tbh]
	\begin{center}
		\includegraphics{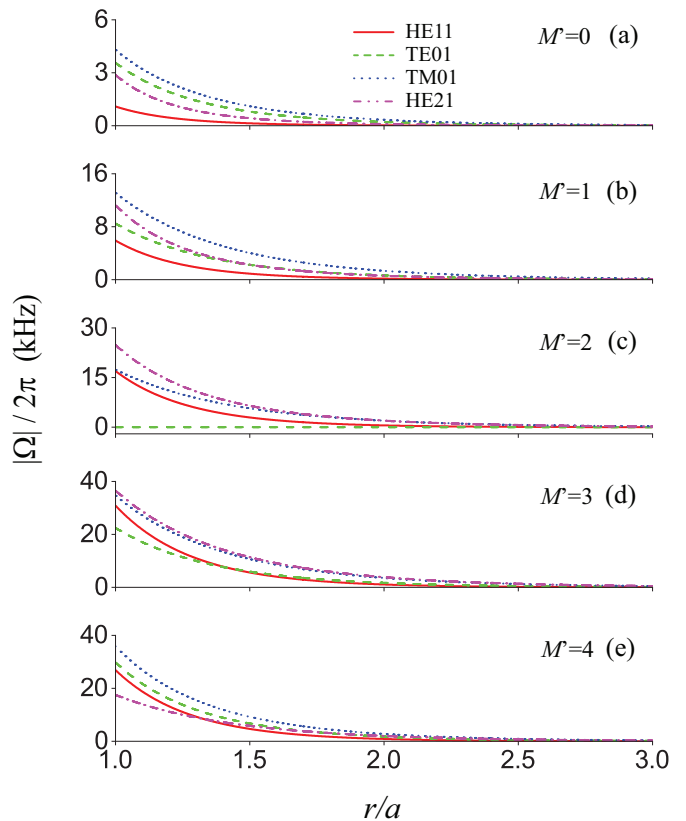}
	\end{center}
	\caption{Absolute value of the Rabi frequency $\Omega$ for the quadrupole transition between the sublevel $M=2$ of the hfs level $5S_{1/2}F=2$ and a sublevel $M'$ of the hfs level $4D_{5/2}F'=4$ of a $^{87}$Rb atom as a function of the radial distance $r$ for different magnetic quantum numbers $M'=0$, 1, 2, 3, and 4 and different guided mode types HE$_{11}$, TE$_{01}$, TM$_{01}$, and HE$_{21}$. The fiber radius is $a=280$ nm. The wavelength of the atomic transition is $\lambda_0=516.5$ nm. The refractive indices of the fiber and the vacuum cladding are $n_1=1.4615$ and $n_2=1$, respectively. The power of the guided light field is $1$ nW. The field propagates in the $+z$ direction, and the hybrid modes HE$_{11}$ and HE$_{21}$ are counterclockwise quasicircularly polarized. The quantization axis is $x_3\parallel z$.}
	\label{fig2}
\end{figure}

The axial torque $T_z$ depends on the Rabi frequency $\Omega$.
We plot in Fig.~\ref{fig2} the absolute value $|\Omega|$ of the Rabi frequency for the quadrupole transition between 
the sublevel $M=2$ of the hyperfine (hfs) level $5S_{1/2}F=2$ and a sublevel $M'$ of the hfs level $4D_{5/2}F'=4$
as a function of the radial distance $r$ for different magnetic quantum numbers $M'=0$, 1, 2, 3, and 4 and different guided mode types HE$_{11}$, TE$_{01}$, TM$_{01}$, and HE$_{21}$.
Figure \ref{fig2} shows that $|\Omega|$ decreases almost exponentially with increasing radial distance $r$. The steep slope in the radial dependence of $|\Omega|$ is a consequence of the evanescent-wave behavior of the guided field outside the fiber.
We observe from Fig.~\ref{fig2} that $|\Omega|$ depends on the type of the guided mode and
the magnetic quantum numbers of the atomic transition. 
Note that the dashed green curve in Fig.~\ref{fig2}(c), which stands for the case of the atom with the levels $M'=M=2$ interacting with the field in the TE mode, is zero \cite{quadrupole}. This is a consequence of the specific properties of the TE mode and the quadrupole operator $Q_{ij}$ for the transition $|F=2, M=2\rangle\to |F'=4, M'=2\rangle$ with the quantization axis $x_3\parallel z$.

\begin{figure}[tbh]
	\begin{center}
		\includegraphics{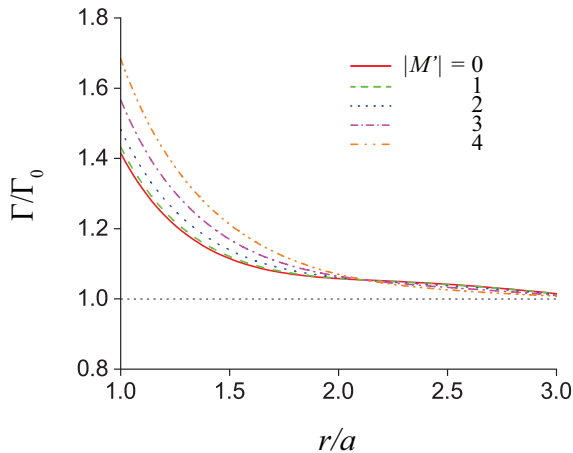}
	\end{center}
	\caption{
		Dipole decay rate $\Gamma$ for a  magnetic sublevel $M'$ of the hfs level $4D_{5/2}F'=4$ of a $^{87}$Rb atom
		as a function of the radial distance $r$ for different quantum numbers $|M'|=0$, 1, 2, 3, and 4.
		The fiber radius is $a=280$ nm. The wavelength of the dipole transition between the levels $4D_{5/2}$ and $5P_{3/2}$ is $1528.95$ nm. 
		The corresponding refractive index of the fiber is $n_1=1.4443$.
		The rate is normalized to the free-space atomic decay rate $\Gamma_0=1.119\times10^7$ s$^{-1}$. 
	}
	\label{fig3}
\end{figure}

The axial torque $T_z$ also depends on the decay rate $\Gamma$ of the excited state.
For the level $4D_{5/2}$ of atomic rubidium, $\Gamma$ is mainly determined by the dipole transition from this level to the level $5P_{3/2}$ with the wavelength $1528.95$ nm \cite{NIST}. When the atom is in free space, the decay rate is $\Gamma=\Gamma_0=1.119\times10^7$ s$^{-1}$ \cite{Safronova2011}.  When the atom is in the vicinity of a nanofiber, $\Gamma$ is modified \cite{cesium decay}. We use the technique of Ref.~\cite{cesium decay} to calculate $\Gamma$.
We plot in Fig.~\ref{fig3} the radial dependencies of  $\Gamma$ for different magnetic sublevels $M'$ of the hfs level $4D_{5/2}F'=4$. The figure shows that $\Gamma$ is enhanced and depends on the magnetic sublevel $M'$ and the radial distance $r$. It is interesting to note that all the curves for different $M'$ cross each other at a radial point $r/a\cong 2.12$. The reason is that at this point the fiber-modified decay rates for the $\sigma^{\pm}$ and $\pi$ transitions are equal to each other and, hence, the decay rate of the magnetic sublevel $M'$ does not depend on $M'$.

\begin{figure}[tbh]
	\begin{center}
		\includegraphics{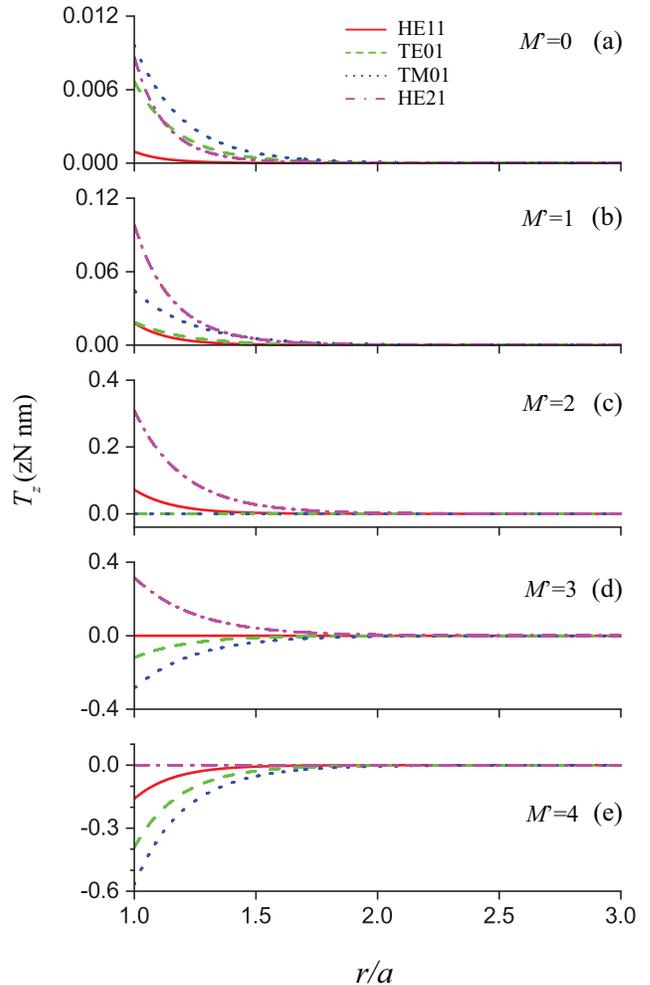}
	\end{center}
	\caption{Torque $T_z$ produced by the quadrupole transition between the sublevel $M=2$ of the hfs level $5S_{1/2}F=2$ and a sublevel $M'$ of the hfs level $4D_{5/2}F'=4$ of a $^{87}$Rb atom as a function of the radial distance $r$ for different magnetic quantum numbers $M'=0$, 1, 2, 3, and 4 and different guided mode types $N=$ HE$_{11}$, TE$_{01}$, TM$_{01}$, and HE$_{21}$. The field detuning is $\Delta=0$. Other parameters are as for Figs.~\ref{fig2} and \ref{fig3}.}
	\label{fig4}
\end{figure}

We use Eq.~(\ref{a24}) to calculate the torque $T_z$ produced by the quadrupole transition between 
the sublevel $M=2$ of the hfs level $5S_{1/2}F=2$ and a sublevel $M'$ of the hfs level $4D_{5/2}F'=4$ of a $^{87}$Rb atom. We plot in Fig.~\ref{fig4} the torque $T_z$ as a function of the radial distance $r$ for different magnetic quantum numbers $M'=0$, 1, 2, 3, and 4 and different guided mode types HE$_{11}$, TE$_{01}$, TM$_{01}$, and HE$_{21}$. We observe that the torque depends on the atomic transition and the field mode.
The dashed green and dotted blue curves in  Fig.~\ref{fig4}(c) show that $T_z$ is vanishing for the TE and TM modes (with $l=0$) in the case  $M'-M=0$. Similarly, the solid red curve in Fig.~\ref{fig4}(d) and the dash-dotted magenta curve in Fig.~\ref{fig4}(e) indicate that $T_z=0$ for the HE$_{11}$ mode
(with $l=1$) in the case $M'-M=1$ and for the HE$_{21}$ mode (with $l=2$) in the case $M'-M=2$. Such a vanishing of the  torque $T_z$ occurs when $pl=M'-M$ [see Eq.~(\ref{a24})], that is, when the angular momentum per photon is equal to the change in the angular momentum of the atomic internal state per transition. When $pl\not=M'-M$, the torque $T_z$ is nonzero and is governed by the conservation law expression (\ref{b7}). It is interesting to note from Fig.~\ref{fig4} that the sign of $T_z$ can be positive or negative depending on the sign of the factor $pl-M'+M$ [see Eq.~(\ref{a24})]. Comparison between the solid red and dash-dotted magenta curves of Fig.~\ref{fig4} shows that the absolute value of the torque for the higher-order mode HE$_{21}$ (see the dash-dotted magenta curves) is larger than that of the torque for the fundamental mode HE$_{11}$ (see the solid red curves) except for the case of Fig.~\ref{fig4}(e), where the torque for the mode HE$_{21}$ is vanishing for $pl=M'-M=2$.

We note that the maximal values of the axial torque $T_z$ in Fig.~\ref{fig4} are on the order of $0.6$ zN nm [see the dotted blue curve in Fig.~\ref{fig4}(e)]. The power of 1 nW for the driving guided field is used in our numerical calculations. For the radial distance $r=300$ nm from the fiber center, the corresponding azimuthal force $F_\varphi$ is on the order of $0.002$ zN. Such an optical quadrupole force is significantly weaker than the typical dipole forces ($\sim10$ zN) on single atoms in laser cooling and trapping techniques \cite{coolingbook}. By increasing the power of the guided driving field, we can achieve larger forces and torques on atoms with quadrupole transitions.
Note that the power of a few $\mu$W for the driving guided field was used in the experiment \cite{Ray2020}. For such a power, an azimuthal force on the order of 1 zN and an axial torque on the order of $1000$ zN nm can be achieved.

We do not calculate numerically the torque of scattering (re-emission) of light from an atom in a magnetic sublevel $M'$ of the hfs state $4D_{5/2}F'=4$. The scattering from this state is mainly determined by the dipole transition between it and the state $5P_{3/2}F=3$. The numerical calculations for the torque produced by this scattering process would involve the multilevel structure of the atom and are beyond the scope of this paper. In the framework of the model of a two-level atom with a dipole transition, the torque of scattering of nanofiber-guided light has been studied analytically and numerically \cite{chiraltorque}.

\section{Summary}
\label{sec:summary}

In conclusion, we have studied the transfer of angular momentum of guided photons to
a two-level atom with an electric quadrupole transition near an optical nanofiber. 
We have shown that the generation of the axial orbital torque of the driving guided field on the atom is governed by the internal-state selection rules for the quadrupole transition and by the angular momentum conservation law with the photon angular momentum given in the Minkowski formulation. We have found that the torque depends on the photon angular momentum, the change in the angular momentum of the atomic internal state, and the quadrupole-transition Rabi frequency. We have calculated numerically the torques for the quadrupole transitions between the sublevel $M=2$ of the hfs level $5S_{1/2}F=2$ and the  sublevels $M'=0$, 1, 2, 3, and 4 of the hfs level $4D_{5/2}F'=4$ of  a $^{87}$Rb atom. We have shown that the absolute value of the torque for the higher-order mode HE$_{21}$  is larger than that of the torque for the fundamental mode HE$_{11}$ except for the case $M'-M=2$, where the torque for the mode HE$_{21}$ is vanishing. 
Our results are important for generation, control, and manipulation of orbital angular momenta of atoms using nanofiber guided light.

\begin{acknowledgments}
	
The authors are thankful to M. Babiker and J. Everett for useful discussions.
The authors acknowledge the supports from the Okinawa Institute of Science and Technology (OIST) Graduate University and from the Japan Society for the Promotion of Science (JSPS) Grant-in-Aid for Scientific Research (C) under Grants No. 19K05316 and No. 20K03795.
\end{acknowledgments}


\end{document}